\documentclass[twocolumn,secnumarabic,amssymb,nobibnotes,showpacs,aps,prd]{revtex4-1}

\usepackage{hyperref}
\usepackage{graphicx}
\usepackage{amsmath}

\begin{document}

\title{Mass spectra and decay properties of $D_s$ Meson  in a relativistic Dirac formalism }

\author{Manan Shah$^1$}
\email[]{mnshah09@gmail.com}

\author{Bhavin Patel$^2$}
\email[]{azadpatel2003@yahoo.co.in}
\author{P C Vinodkumar$^1$}
\email{p.c.vinodkumar@gmail.com}
\affiliation{$^1$Department of Physics, Sardar Patel University,Vallabh Vidyanagar, INDIA.}
\affiliation{$^2$P. D. Patel Institute of Applied Sciences, CHARUSAT, Changa, INDIA}
\date{\today}%

\begin{abstract}
The mass spectra of $D_s$ meson is obtained in the framework of relativistic independent quark model using Martin like potential for the quark confinement.  The predicted excited states are in good agreement with the experimental results as well as with the lattice and other theoretical predictions. The spectroscopic parameters are employed further to compute the decay constant, electromagnetic transition and leptonic decay widths. The present result for its decay constant, $f_P$ (252.82 MeV) is in excellent agreement with the value 252.6 $\pm$ 11.1 MeV reported by CLEO-c and the predicted branching ratios for $(D_s \rightarrow \tau \bar{\nu}_\tau, \mu \bar{\nu}_\mu)$ ($5.706 \times 10^{-2}, 5.812 \times 10^{-3}$) are in close agreement with the PDG values ($ (5.43 \pm 0.31)\times 10^{-2}, (5.90 \pm 0.33)\times 10^{-3} $) respectively.
\end{abstract}
\maketitle

\section{Introduction} \label{intro}
Having played a major role in the foundation of QCD, heavy hadron spectroscopy has witnessed in the last few years a renewal of interest due to many new states observed in recent years. The remarkable progress at the experimental \cite{PDG2012} side, with various high energy machines such as BaBar, BELLE, BES-III, B$-$factories, Tevatron, ARGUS collaborations, CLEO, CDF, SELEX, D${\O}$ etc., for the study of hadrons has opened up new challenges in the theoretical understanding of light-heavy flavour hadrons. Study of $D_s$ meson carry special interest as it is a hadron with two open flavours ($c, \bar s $) that restricts its decay via strong interactions. These particles thus provide us a clean laboratory to study electromagnetic and weak interaction. The discoveries of new resonances of $D_s$ states such as $D_s$(2638) \cite{Evdokimov}, $D_s$(2710) \cite{Brodzicka}, $D_s$(2860) \cite{Aubert09}, $D_s$(3040) \cite{Aubert09} etc., have further generated considerable interest towards the spectroscopy of this double open flavour meson.  The masses of low-lying 1S and $1P_J$ states of $D_s$ mesons are recorded both experimentally \cite{PDG2012} and theoretically \cite{Godfrey,Pierro,Ebert,Bardeen,Colangelo,Falk,Eichen,Bhavin2010,Naynesh2011}. However, the existing results on excited heavy-light mesons are partially inconclusive and even contradictory in several cases.\\

Thus any attempts towards the understanding of these newly observed states become very important for better understanding the quark-antiquark dynamics within $Q\bar{q}$ bound state. So, a successful theoretical model can provide important information about the quark-antiquark interactions and the behavior of QCD within the doubly open flavour at the hadronic system. Though there exist many theoretical models to study the hadron properties based on its quark structure,  the predictions for low-lying states are off by $60-90$ MeV with respect to the respective experimental values. Moreover the issue related to the hyperfine and fine structure splitting of the mesonic states; their intricate dependence with the constituent quark masses and the running strong coupling constant are still unresolved. Though the validity of nonrelativistic models is very well established and significantly successful for the description of heavy quarkonia, it seemed to fail for the description of meson containing light flavour quarks or antiquarks. \\

Apart from the successful predictions of the mass spectra, validity of any phenomenological model  depends also on the successful predictions of their decay properties. For better predictions of the decay widths, many models have incorporated additional contributions such as radiative and higher order QCD corrections \cite{Bhavin2010, Rai2008,Ebert03,Lansberg,Kim}. Thus, in this paper we make an attempt to study properties like mass spectrum, decay constants and other decay properties of the $D_s$ meson based on a relativistic Dirac formalism. We investigate the heavy-light mass spectra of $D_s$($c\bar{s}$) meson in this framework with Martin like confinement potential. \\

Along with the mass spectra, the pseudoscalar decay constants of the heavy-light mesons have also been estimated in the context of many QCD-motivated approximations. The predictions of such methods cover a wide range of values \cite{Wang,Cvetic}. It is important to have reliable estimate of the decay constant as it is an important parameter in many weak processes such as quark mixing, CP violation, etc. The leptonic decay of charged meson is another important annihilation channel through the exchange of virtual $W^{\pm}$ boson. Though this annihilation process is rare, but they have clear experimental signatures due to the presence of highly energetic leptons in the final state. And there exist experimental observations of the leptonic decays of $D_s$ meson. The leptonic decays of mesons entails an appropriate representation of the initial state of the decaying vector
mesons in terms of the constituent quark and antiquark with their respective momenta and spin. The bound constituent quark and antiquark inside the
meson are in definite energy states having no definite momenta. However one can find out the momentum distribution amplitude for the constituent quark
and antiquark inside the meson immediately before their annihilation to a lepton pair. Thus, it is appropriate to compute the leptonic branching ratio here and compare our result with the experimental values as well as with the predictions based on other models.
\section{Theoretical Framework}
 The quark confining interaction of meson is considered to be produced by the non-perturbative multigluon mechanism and this mechanism is unfeasible to estimate theoretically from first principles of QCD. In the present study, we assume that the constituent quarks inside a meson is independently confined by an average potential of the form \cite{Barik2000}
\begin{equation}\label{eq:a}
V(r)= \frac{1}{2} (1+\gamma_0) (\lambda r^{0.1}+V_0)
\end{equation}
To first approximation, the confining part of the interaction is believed to provide the zeroth-order quark dynamics inside the meson through the quark Lagrangian density

\begin{equation}\label{eq:b}
{\cal L}^0_q (x)= \bar{\psi}_q(x) \left[\frac{i}{2} \gamma^{\mu} \overrightarrow{\partial_\mu} - V(r) - m_q \right] \psi_q(x).
\end{equation}
In the stationary case, the spatial part of the quark wave functions $\psi(\vec{r})$ satisfies the Dirac equation given by
\begin{equation}\label{eq:c}
[\gamma^0 E_q - \vec{\gamma}. \vec{P} - m_q - V (r)]\psi_q (\vec{r}) = 0.
\end{equation}

The solution of Dirac equation can be written as two component (positive and negative energies in the zeroth order) form as
\begin{equation}\label{eq:d}
\psi_{nlj}(r) = \left(
    \begin{array}{c}
      \psi_{nlj}^{(+)} \\
      \psi_{nlj}^{(-)}
    \end{array}
  \right)
\end{equation}
where
\begin{equation}\label{eq:e}
\psi_{nlj}^{(+)}(\vec{r}) = N_{nlj} \left(
    \begin{array}{c}
      i g(r)/r \\
      (\sigma.\hat{r}) f(r)/r
    \end{array}
  \right) {\cal{Y}}_{ljm}(\hat{r})
\end{equation}
\begin{equation}\label{eq:f}
\psi_{nlj}^{(-)}(\vec{r}) = N_{nlj} \left(
    \begin{array}{c}
      i (\sigma.\hat{r}) f(r)/r \\
        g(r)/r
    \end{array}
  \right) (-1)^{j+m_j-l} {\cal{Y}}_{ljm}(\hat{r})
\end{equation}\\
and $N_{nlj}$ is the overall normalization constant. The normalized spin angular part is expressed as
\begin{equation}\label{eq:g}
{\cal{Y}}_{ljm}(\hat{r}) = \sum_{m_l, m_s}\langle l, m_l, \frac{1}{2}, m_s| j, m_j \rangle Y^{m_l}_l \chi^{m_s}_{\frac{1}{2}}
\end{equation}
Here the spinor $\chi_{\frac{1}{2}{m_s}}$ are eigenfunctions of the spin operators,
\begin{equation}\label{eq:h}
\chi_{\frac{1}{2} \frac{1}{2}} = \left(
    \begin{array}{c}
      1 \\
      0
    \end{array}
  \right) \ \ \ , \ \ \ \ \chi_{\frac{1}{2} -\frac{1}{2}} = \left(
    \begin{array}{c}
      0 \\
      1
    \end{array}
  \right)
\end{equation}
The reduced radial part $g(r)$ of the upper component and $f(r)$ of the lower component of Dirac spinor $\psi_{nlj}(r)$ are the solutions of the equations given by\\
\begin{equation}\label{eq:i1}
\frac{d^2 g(r)}{dr^2}+\left[(E_{D}+ m_q) [E_{D} - m_q - V(r)] - \frac{\kappa (\kappa + 1)}{r^2}\right] g(r) = 0
\end{equation}
and
\begin{equation}\label{eq:i2}
\frac{d^2 f(r)}{dr^2}+\left[(E_{D}+ m_q) [E_{D} - m_q - V(r)] - \frac{\kappa (\kappa - 1)}{r^2}\right] f(r) = 0
\end{equation}
It can be transformed in to a convenient dimensionless form given as \cite{barik1982}
\begin{equation}\label{eq:j1}
\frac{d^2 g(\rho)}{d\rho^2}+\left[\epsilon- \rho^{0.1} - \frac{\kappa (\kappa+1)}{\rho^2}\right] g(\rho) = 0
\end{equation}
 and
\begin{equation}\label{eq:j2}
\frac{d^2 f(\rho)}{d\rho^2}+\left[\epsilon- \rho^{0.1} - \frac{\kappa (\kappa -1)}{\rho^2}\right] f(\rho) = 0
\end{equation}
In terms of  dimensionless variable $\rho = (r / r_0)$ with the arbitrary scale factor chosen conveniently as
\begin{equation}\label{eq:k}
r_0 = \left[(m_q + E_{D})\frac{\lambda}{2}\right]^{-\frac{10}{21}},
\end{equation}
and a corresponding dimensionless energy eigenvalue defined as
\begin{equation}\label{eq:l}
\epsilon = (E_{D} - m_q - V_0) (m_q + E_{D})^{\frac{1}{21}} \left(\frac{2}{\lambda}\right)^{\frac{20}{21}}
\end{equation}
Here, it is suitable to define a quantum number $\kappa $ by
\begin{eqnarray}\label{eq:kappa}
\kappa = &\left\{\begin{matrix} -(\ell + 1)& = - \left(j+\frac{1}{2}\right) & \ \ for \ \ j= \ell+\frac{1}{2}\\
                                      \ell & = + \left(j+\frac{1}{2}\right) & \ \ for \ \ j= \ell-\frac{1}{2} \end{matrix}\right.
\end{eqnarray}
Equations (\ref{eq:j1}) and (\ref{eq:j2}) now can be treated similar to radial Schr$\ddot{o}$dinger equation with a potential $\rho^{\nu}$ which can be solved numerically \cite{Bhavin2009JPG}.


The solutions $g (\rho)$ and $f (\rho)$ are normalized to get
\begin{equation}
 \int_0^\infty (f_q^2(r) + g_q^2(r)) \ d r = 1.
\end{equation}

\begin{table*}
\begin{center}
\caption{The fitted model parameters for the $D_s$ systems}\label{parameter}
\begin{tabular}{|c|c|}
\hline
\textbf{System Parameters } & \textbf{$D_s$}  \\
\hline\hline
Quark mass (in GeV)& $m_s = $ 0.1 and $m_c$ = 1.27 	   \\

\hline
Potential parameter($\lambda$) & $2.2655 + B $ $GeV^{\nu+1}$ \\
\hline
$V_0$     &- 2.6155 GeV    \\
\hline
Centrifugal parameter (B)  & $(n*0.153)\ GeV^{-1}$  for $l = 0$   \\
                           & $((n+l)*0.1267)\ GeV^{-1}$ for $l \neq 0$  \\
\hline
$\sigma$ ($j-j$ coupling constant) & $0.0055 \ GeV^3$ for $l = 0$ \\
                                & $0.2696 \ GeV^3$ for $l \neq 0$ \\
\hline
\end{tabular}
\end{center}
\end{table*}



The wavefunction for a $D_s (c\bar s)$  meson now can be constructed using Eqn (\ref{eq:e}) and (\ref{eq:f}) and the corresponding mass of the quark-antiquark system can be written as
\begin{equation}
M_{Q \bar{q}} = E_{D}^Q + E_{D}^{\bar{q}}
\end{equation}
where $E_D^{Q/\bar{q}}$ are obtained using Eqn. (\ref{eq:l}) and (\ref{eq:kappa}). For the spin triplet (vector) and spin singlet (pseudoscalar) state, the choices of ($j_1$, $j_2$) are $\left(\left(l_1 + \frac{1}{2}\right), \left(l_1 + \frac{1}{2}\right)\right)$ and $\left(\left(l_1 + \frac{1}{2}\right), \left(l_1 - \frac{1}{2}\right)\right)$ respectively. The previous work of independent quark model within the Dirac formalism by \cite{Barik2000} has been extended here by incorporating the spin-spin, spin-orbit and tensor interactions of the confined one gluon exchange potential (COGEP) \cite{PCV1992,Khadkikar1991}, in addition to the j-j coupling of the quark-antiquark. Finally, the mass of the specific $^{2 S+1}L_J$ states of $Q \bar q$ system is expressed as
\begin{eqnarray}
M_{^{2 S+1}L_J} =  M_{Q \bar q} \ (n_1l_1j_1, n_2l_2j_2) &&+  \langle V_{Q \bar q}^{j_1j_2}\rangle  \\ && + \langle V_{Q \bar q}^{LS}\rangle + \langle V_{Q \bar q}^{T}\rangle \nonumber
\end{eqnarray}
Here, the spin-spin  part is defined as
\begin{equation}
\langle V^{j_1 j_2}_{Q \bar q} (r)\rangle = \frac{\sigma \ \langle j_1 j_2 J M |\hat{j_1}.\hat{j_2}| j_1 j_2 J M \rangle}{(E_Q + m_{Q})(E_{\bar{q}} + m_{\bar{q}})}
\end{equation}
where $\sigma$ is the $j-j$ coupling constant. The expectation value of $\langle j_1 j_2 J M |\hat{j_1}.\hat{j_2}| j_1 j_2 J M \rangle$ contains the ($j_1.j_2$) coupling and the square of Clebsch-Gordan coefficients. The tensor and spin-orbit parts of confined one-gluon exchange potential (COGEP) \cite{PCV1992,Khadkikar1991} is given by
\begin{eqnarray}\label{Vt}
V^{T}_{Q \bar q} (r) &=& - \frac{\alpha_s}{4} \frac{N_Q^2 N_{\bar q}^2}{(E_Q + m_{Q})(E_{\bar{q}} + m_{\bar{q}})} \nonumber\\ &&\otimes \ \lambda_Q . \lambda_{\bar q} \left( \left( \frac{D''_1 (r)}{3}- \frac{D'_1 (r)}{3 \ r} \right) S_{Q \bar q}\right)
\end{eqnarray}
where $S_{Q \bar q} = \left[ 3 (\sigma_Q. \vec{\hat{r}})(\sigma_{\bar q}. \vec{\hat{r}})- \sigma_Q . \sigma_{\bar q}\right]$ and $\vec{\hat{r}} = \vec{\hat{r}}_Q - \vec{\hat{r}}_{\bar q}$ is the unit vector in the direction of $\vec{r}$ and
\begin{eqnarray}\label{Vls}
V^{LS}_{Q \bar q} (r) &=& \frac{\alpha_s}{4} \frac{N_Q^2 N_{\bar q}^2}{(E_Q + m_{Q})(E_{\bar{q}} + m_{\bar{q}})}  \frac{\lambda_Q . \lambda_{\bar q}}{2 \ r} \\ &&\otimes \left[ \left[ r \times (\hat{p_Q}-\hat{p_q}). (\sigma_Q + \sigma_q)\right]\left( {D'_0 (r)}+ 2 {D'_1 (r)} \right) \right. \nonumber\\ &&  \left. + \left[ r \times (\hat{p_Q}+\hat{p_q}). (\sigma_i - \sigma_j)\right]\left( {D'_0 (r)}-  {D'_1 (r)} \right) \right]  \nonumber
\end{eqnarray}
where $\alpha_s$ is the strong coupling constant and it is computed as
 \begin{equation}
 \alpha_s = \frac{4 \pi}{(11-\frac{2}{3}\  n_\emph{f})\log\left(\frac{E^2_Q}{\Lambda^2_{QCD}}\right)}
 \end{equation}
 with $n_\emph{f}$ = 3 and $\Lambda_{QCD}$ = 0.150 GeV. In Eqn. (\ref{Vls}) the spin-orbit term has been split into symmetric $(\sigma_Q + \sigma_q)$ and anti-symmetric $(\sigma_Q - \sigma_q)$ spin-orbit terms.

We have adopted the same parametric form of the confined gluon propagators which are given by \cite{PCV1992,Khadkikar1991}
\begin{equation}
D_0 (r) = \left( \frac{\alpha_1}{r}+\alpha_2 \right) \exp(-r^2 c_0^2/2)
\end{equation}
and
\begin{equation}
D_1 (r) =  \frac{\gamma}{r} \exp(-r^2 c_1^2/2)
\end{equation}
with $\alpha_1$ = 0.036, $\alpha_2$ = 0.056, $c_0$ = 0.1017 GeV, $c_1$ = 0.1522 GeV, $\gamma$ = 0.0139. Other optimized model parameters employed in the present study are listed in the Table \ref{parameter}. The computed S-wave masses and other P-wave and D-wave masses of $D_s$ meson states are listed in Table \ref{tab1} and Table \ref{tab2} respectively. Fig.(\ref{mass spectra}) shows the energy level diagram of $D_s$ meson spectra along with available experimental results.\\

\begin{table*}
\begin{center}
\caption{S-wave $D_s$ ($c\bar{s}$) spectrum (in MeV).} \label{tab1}
\begin{tabular}{cccccccccccccc}
\hline\hline
 &  &    &  &       &&    & \multicolumn{2}{c} {Experiment} &&&& &\\
\cline{8-9}
nL && $J^P$ & State &$M_{Q \bar q}$& $\langle V_{Q \bar q}^{j_1j_2}\rangle$ &Present  &  Meson  & Mass\cite{PDG2012} && \cite{Badalian2011} & \cite{Ebert2010} & \cite{Naynesh2011} & \cite{De2011}\\
\hline\\
1S && $1^-$  &$1{^3S_1}$& 2113.2 & 0.73 &2113.9 & $D_s^*$   &  2112.3 $\pm$ 0.5   && . . . & 2111 & 2117 & 2107\\
   && $0^-$  &$1{^1S_0}$& 1970.1 &-1.84 &1968.3 & $D_s$     &  1968.49 $\pm$ 0.32 && . . . & 1969 & 1970 & 1969\\\\

2S && $1^-$  &$2{^3S_1}$& 2717.3 &0.46  &2717.8 &  $D^*_{s}$(2710) &  $2710_{-7}^{+12}$ \cite{Brodzicka2008,Aubert2006}&& 2728 & 2731 & 2723 & 2714\\
   && $0^-$  &$2{^1S_0}$& 2634.6 &-1.06 &2633.5 &  $D_s$(2632)     &  2632.5 $\pm$ 1.7 \cite{Evdokimov}  && 2656 & 2688 & 2684 & 2640\\\\

3S && $1^-$  &$3{^3S_1}$& 3263.5 &0.33  &3263.8 &        &  . . . && 3200 & 3242 & 3180 &. . .\\
   && $0^-$  &$3{^1S_0}$& 3203.2 &-0.75 &3202.4 &        &  . . . && 3140 & 3219 & 3158 &. . .\\\\

4S && $1^-$  &$4{^3S_1}$& 3781.4 &0.25  &3781.6 &        &  . . . && . . . & 3669 & 3571  &. . .\\
   && $0^-$  &$4{^1S_0}$& 3732.7 &-0.57 &3732.1 &        &  . . .  && . . . & 3652 & 3556 &. . .\\\\
\hline\hline

\end{tabular}

\end{center}
\begin{center}
    \cite{Badalian2011} - Semi-relativistic model\\ \cite{Ebert2010} - Quasi potential Approach \\ \cite{Naynesh2011} - Relativistic quark-antiquark potential (Coulomb plus power)
model\\ \cite{De2011} - Non-relativistic constituent quark model
\end{center}

\end{table*}

\begin{table*}
\begin{center}
\caption{P-wave and D-wave $D_s$ ($c\bar{s}$) spectrum (in MeV).} \label{tab2}
\begin{tabular}{cccccccccccccc}
\hline\hline
 &  &   &   &&&&           & \multicolumn{2}{c} {Experiment} & &&&\\
\cline{9-10}
nL & $J^P$ & State &$M_{Q \bar q}$ &$\langle V_{Q \bar q}^{j_1j_2}\rangle$ &$ \langle V^{T}\rangle$& $\langle V^{LS}\rangle$& Present  &  Meson  & Mass \cite{PDG2012} & \cite{Badalian2011} & \cite{Ebert2010}& \cite{Naynesh2011} & \cite{De2011}\\
\hline\\
1P & $2^+$  &$1{^3P_2}$& 2520.9 &19.24  &-3.71  &48.23  &2584.7 & $D_{s2}$(2573)   &  2571.9 $\pm$ 0.8   & . . . & 2571 &2566 & 2559 \\
   & $1^+$  &$1{^3P_1}$& 2520.9 &25.65  &18.54  &-48.23 &2516.9 & $D_{s1}$(2536)   &  2535.12 $\pm$ 0.13 & . . . & 2536 &2540 & 2510\\
   & $0^+$  &$1{^3P_0}$& 2520.9 &-38.47 &-37.08 &-96.46 &2349.0 & $D_{s0}$(2317)   &  2317.8 $\pm$ 0.6   & . . . & 2509 &2444 & 2344\\
   & $1^+$  &$1{^1P_1}$& 2421.7 &13.84  &0      & 0     &2435.6 &  $D_{s1}$(2460)   &  2459.6 $\pm$ 0.6  & . . . & 2574 &2530 & 2488\\\\

2P & $2^+$   &$2{^3P_2}$& 3018.3 &13.87  &-6.28 &81.75   &3107.6 &                  &                      & 3045 & 3142 & 3048 & 3040\\
   & $1^+$   &$2{^3P_1}$& 3018.3 &18.50  &31.40 &-81.75  &2986.4 & $D_{sJ}$(3040)  &  $3044_{-9}^{+30}$ \cite{Aubert09}& 3040 & 3067 & 3019& 2958\\
   & $0^+$   &$2{^3P_0}$& 3018.3 &-27.75 &-62.8 &-163.51 &2764.3 &                 &                       & 2970 & 3054 & 2947 & 2830 \\
   & $1^+$   &$2{^1P_1}$& 2949.4 & 9.64  &  0   &  0     &2959.0 &                 &                       & 3020 & 3154 & 3023 & 2995\\\\

3P & $2^+$  &$3{^3P_2}$& 3479.7 &10.76 &-8.53 &111.06  &3593.0 &                  &         & . . . & 3580 & . . . & . . . \\
   & $1^+$  &$3{^3P_1}$& 3479.7 &14.34 &42.64 &-111.06 &3425.6 &                  &         & . . . & 3519 & . . . & . . . \\
   & $0^+$  &$3{^3P_0}$& 3479.7 &-21.51&-85.27&-222.13 &3150.9 &                  &         & . . . & 3513 & . . . & . . . \\
   & $1^+$  &$3{^1P_1}$& 3426.0 &7.37  &  0  & 0       &3433.4 &                  &         & . . . & 3618 & . . . & . . . \\
\hline\\
1D & $3^-$  &$1{^3D_3}$& 2952.7 &-21.79 &-0.03 &0.49 & 2931.4 & $D^*_{sJ}$(2860)&  $2862_{-3}^{+6}$ \cite{Aubert09}& 2840 & 2971 & 2834& 2811\\
   & $2^-$  &$1{^3D_2}$& 2952.7 &-64.74 & 0.11 &-0.25& 2887.8 &                  &                   & 2885 & 2961 & 2816 & 2788\\
   & $1^-$  &$1{^3D_1}$& 2952.7 &-109.81&-0.11 &-0.75& 2842.0 &                  &                   & 2870 & 2913 & 2873 & 2804\\
   & $2^-$  &$1{^1D_2}$& 2874.3 &-2.65  & 0    & 0   & 2871.6 &                  &                   & 2828 & 2931 & 2896 & 2849\\\\

2D & $3^-$  &$2{^3D_3}$& 3423.7 &-15.72 &-0.03 &0.52 & 3408.4 &                  &                   & 3285 & 3469 & 3263 & 3240\\
   & $2^-$  &$2{^3D_2}$& 3423.7 &-46.71 & 0.11 &-0.26& 3376.8 &                  &                   & . . .& 3456 & 3248 & 3217\\
   & $1^-$  &$2{^3D_1}$& 3423.7 &-79.23 &-0.11 &-0.79& 3343.5 &                  &                   & 3290 & 3383 & 3292 & 3217\\
   & $2^-$  &$2{^1D_2}$& 3363.7 &-1.87  & 0    & 0   & 3361.8 &                  &                   & . . .& 3403 & 3312 & 3260\\\\

3D & $3^-$  &$3{^3D_3}$& 3870.9&-12.06 &-0.04&0.60 & 3859.4 &                  &                   & . . . & . . . &. . . &. . .\\
   & $2^-$  &$3{^3D_2}$& 3870.9&-35.85 & 0.13&-0.30& 3834.9 &                  &                   & . . . & . . . & . . .&. . .\\
   & $1^-$  &$3{^3D_1}$& 3870.9&-60.80 &-0.13&-0.91& 3809.1 &                  &                   & . . . & . . . & . . .&. . .\\
   & $2^-$  &$3{^1D_2}$& 3821.7&-1.42  & 0   & 0   & 3820.3 &                  &                   & . . . & . . . &. . . &. . .\\\\
   \hline\hline
\end{tabular}
\end{center}
\end{table*}
\section{ Magnetic (M1) Transitions of Open Charm Meson}
Spectroscopic studies led us to compute the decay widths of energetically allowed radiative transitions of the type $A \rightarrow B + \gamma $  among several vector and pseudoscalar states of $D_s$ meson. The magnetic transition correspond to spin flip and hence the vector meson decay to pseudoscalar $V\rightarrow P\gamma$ represents a typical M1 transition. Such transitions are experimentally important to the identification of newly observed states. Assuming that such transitions are single vertex processes governed mainly by photon emission from independently confined quark and antiquark inside the meson, the S-matrix elements in the rest frame of the initial meson is written in the form
\begin{equation}\label{eq:q}
S_{BA} = \left<B\gamma \left|-ie\int d^4 x \ T \left[\sum_q e_q \bar{\psi_q} (x) \gamma^\mu \psi_q (x) A_\mu (x) \right] \right|A \right>.
\end{equation}
The common choice of the photon field $A_\mu (x)$ is made here in Coulomb-gauge with $\epsilon (k, \lambda)$ as the polarization vector of the emitted photon having energy momentum $(k_0 = |\textbf{k}|,\textbf{k})$ in the rest frame of A. The quark field operators find a possible expansions in terms of the complete set of positive and negative energy solutions given by Eqs. (\ref{eq:e}) and (\ref{eq:f}) in the form
\begin{eqnarray}\label{eq:r}
\Psi_q (x) = \sum_\zeta &\left[b_{q\zeta} \ \psi_{q\zeta}^{(+)}(r) \ \exp(-iE_{q\zeta}t)  \right. \nonumber \\
  & \left. +\  b_{q\zeta}^{\dag} \ \psi_{q\zeta}^{(-)}(r) \ \exp(iE_{q\zeta}t)\right]
\end{eqnarray}
where the subscript q stands for the quark flavor and $\zeta$ represents the set of Dirac quantum numbers. Here $b_{q\zeta}$ and $b_{q\zeta}^{\dag}$ are the quark annihilation and the antiquark creation operators corresponding to the eigenmodes $\zeta$. After some standard calculations (the details of calculations can be found in Refs. \cite{Barik1992,Barik1993} and \cite{Jena1999}), the S-matrix elements can be expressed as
\begin{eqnarray}\label{eq:s}
S_{BA} &=& i \sqrt{\left(\frac{\alpha}{k}\right)}\  \delta (E_B +k-E_A) \sum_{q,m,m'} \left<B \left| \   \right. \right.  \\
  && \left. \left.  \left[ J^q_{m' m} (k, \lambda) b_{qm'}^\dagger b_{qm} - \  \tilde{J}^{\tilde{q}}_{m m'} (k, \lambda) \tilde{b}_{qm'}^\dagger \tilde{b}_{qm} \right] \right| A \right> \nonumber
\end{eqnarray}
Here $ E_A$ = $M_A$, $ E_B$ = $\sqrt{k^2 + M^2_B}$ and (m, m$'$) are the possible spin quantum numbers of the confined quarks corresponding to the ground state of the mesons. We have
\begin{equation}\label{eq:t}
J^q_{m' m} (k, \lambda) = e_q \int d^3 r \exp (-i \vec k \cdot \vec r) [ \bar{\psi}_{q m'}(r) \vec\gamma  \cdot \vec\epsilon (k, \lambda) \psi_{qm}(r)]
\end{equation}

\begin{equation}\label{eq:u}
\tilde{J}^{\tilde{q}}_{m m'} (k, \lambda)= e_q \int d^3 r \exp (-i \vec k \cdot \vec r) [ \bar{\phi}_{q m}(r) \vec\gamma  \cdot \vec\epsilon (k, \lambda) \phi_{qm'}(r)].
\end{equation}
One can reduce the above equations to simple forms as
\begin{equation}\label{eq:v}
J^q_{m' m} (k, \lambda) = - i\  \mu_q(k)\ [\chi_m^\dagger (\vec \sigma \cdot \vec K) \chi_m ],
\end{equation}
and
\begin{equation}\label{eq:v2}
\tilde{J}^{\tilde{q}}_{m m'} (k, \lambda) = i \ \mu_q(k)\ [\tilde{\chi}_m^\dagger (\vec \sigma \cdot \vec K) \tilde{\chi}_m ]
\end{equation}
where $\vec K = \vec k \times \vec \epsilon (k, \lambda)$. Eqn. (\ref{eq:s}) further simplified to get
\begin{eqnarray}\label{eq:w}
S_{BA} &=& i \sqrt{\left(\frac{\alpha}{k}\right)}\  \delta (E_B +k-E_A) \nonumber \\ && \sum_{q,m,m'} \left<B \left| \mu_q (k) \left[\chi_{m'}^\dagger \vec \sigma \cdot\vec K \chi_m b_{qm'}^\dagger b_{qm}  \right. \right. \right. \nonumber \\ && \left. \left. \left. + \  \tilde{\chi}_m^\dagger \vec \sigma \cdot\vec K \tilde{\chi}_{m'} \tilde{b}_{qm'}^\dagger \tilde{b}_{qm} \right] \right| A \right>
\end{eqnarray}
where $\mu_q(k)$ is expressed as
\begin{equation}\label{eq:x}
\mu_q(k)= \frac{2 e_q}{k} \int_0^\infty j_1(kr)\ f_q (r)\ g_q (r)\ dr
\end{equation}
where $j_1(kr)$ is the spherical Bessel function and the energy of the outgoing photon in the case of a vector meson undergoing a radiative transition to its pseudoscalar state, for instance, $D_s^* \rightarrow D_s \gamma$ is given by
\begin{equation}\label{eq:y}
k = \frac{M_{D_s^*}^2 - M_{D_s}^2}{2 M_{D_s^*}}
\end{equation}
The relevant transition magnetic moment is expressed as
\begin{equation}\label{eq:ab}
\mu_{D_s^* D_s}(k) = \frac{1}{3}[2 \mu_c (k) -  \mu_s (k)],
\end{equation}
Now, the Magnetic (M1) transition width of $D_s^* \rightarrow D_s \gamma$ can be obtained as
\begin{equation}\label{eq:ae}
\Gamma_{D_s^{*} \rightarrow D_s \gamma} = \frac{4 \alpha}{3} k^3 |\mu_{D_s^* D_s} (k)|^2
\end{equation}
The computed transition widths of low lying S-wave states are tabulated in Table \ref{tab9} and are compared with other model predictions.



\section{Decay constant of $D_s$ meson}
The decay constant of a meson is an important parameter in the study of leptonic or non-leptonic
weak decay processes. The decay constant ($f_p$) of pseudoscalar  state is obtained by parameterizing the matrix elements of weak current between the corresponding
meson and the vacuum as \cite{Quang}
\begin{equation}\label{eq:n}
\left<0|\bar{q}\gamma^{\mu} \gamma_5 c| P_\mu \right> = i f_p \ P^\mu
\end{equation}

\begin{table*}
\begin{center}
\caption{Comparison of Center of Mass in $D_s$ meson in MeV.}\label{tab12}
\begin{tabular}{cccccc}
\hline
$M_{CW}$ 	      & &	Present	&	\cite{Mohler2011}	& \cite{Ebert2010} &	 Exp.	  \\
\hline\hline\\				
$\overline{1S}$         && 2077.5	    & $2045.4 \pm 0.215 \pm$ 0.293  &2075.5 &  2076.3  	\\
$\overline{2S}$         && 2696.7 	    &       . . .                      &2720.2 &  2690.6    	\\
$\overline{3S}$         && 3248.4 	    & . . .                            &3236.2 &     . . .      	\\
$\overline{4S}$         && 3769.2 	    &  . . .                           &3664.7 &     . . .        	\\
$\overline{1^3P_J}$     && 2535.9 	    &  . . .                           &2552.4 &     2531.4   	\\
$\overline{1P}$         && 2510.8	    &  . . .                            &2557.8 &  2513.4  \\	
$\overline{2^3P_J}$     && 3029.0 	    &  . . .                            &3107.2 &     . . .   	\\
$\overline{2P}$         && 3011.5 	    &  . . .                            &3118.9 &    . . .      \\	
\hline
\end{tabular}
\end{center}
\end{table*}
\begin{table*}
\begin{center}
\caption{Mass splitting in $D_s$ meson in MeV.}\label{tab11}
\begin{tabular}{cccccc}
\hline
Splitting 	      & &	Present	&	\cite{Mohler2011}	& \cite{Ebert2010} &	 Exp.	  \\
\hline\hline	\\			
$1^3S_1-1 ^1S_0$               &&145.6	    &$133.1 \pm 1.0 \pm$ 1.9   &143  &$143.8 \pm 0.4$  	\\
$2^3S_1-2 ^1S_0$               &&84.3	    &$72 \pm 24 \pm$ 1         &43   &       . . .   	\\
$3^3S_1-3 ^1S_0$               &&61.4 	    &. . .                     &23   &       . . .    	\\
$4^3S_1-4 ^1S_0$               &&49.5 	    & . . .                    &17   &     . . .      	\\
$D_{s0}$(2317)-$\overline{1S}$ &&271.5      &$341.2 \pm 7.7 \pm$ 4.8   &433.5     &241.5 $\pm$ 0.8\\
$D_{s1}$(2460)-$\overline{1S}$ &&358.1      &$459.8 \pm 6.4 \pm$ 6.4   &498.5     & 383.2 $\pm$ 0.8\\
$D_{s1}$(2536)-$\overline{1S}$ &&439.4      &$494.6 \pm 9.2 \pm$ 6.9   &460.5     & 459.0 $\pm$ 0.5\\
$D_{s2}$(2573)-$\overline{1S}$ &&507.2      &$536.7 \pm 9.2 \pm$ 7.5   &495.5     & 496.3 $\pm$ 1.0\\			
$2^1S_0  - \overline{1S}$      &&556.0	    & $654.4\pm 26.7 \pm 9.2$     &612.5       & . . . 	\\
$2^3S_1-\overline{1S}$         &&640.3	    &$726.4\pm 20.8 \pm 10.2$	  &655.5    &$632.7^{+9}_{-6}$    	\\

\hline

\end{tabular}
\end{center}
\end{table*}
It is possible to express the quark-antiquark eigenmodes in the ground state of the meson in terms of the corresponding momentum distribution amplitudes. Accordingly, eigenmodes, $\psi_A^{(+)}$ in the state of definite momentum p and spin projection $s'_p$ can be expressed as
\begin{equation}
\psi_A^{(+)} = \sum_{s'_p} \int d^3 p \ G_q(p,s'_p) \sqrt{\frac{m}{E_p}}\ U_q (p, s'_p) \exp (i\vec{p}\ .\ \vec{r})
\end{equation}
where $U_q (p, s'_p)$ is the usual free Dirac spinors.

In the relativistic quark model, the decay constant can be expressed
through the meson wave function $G_q (p)$ in the momentum
space \cite{Barik1993,HAKAN2000}
\begin{equation}\label{eq:p}
f_P = \left({\frac{3 | I_p |^2}{2 \pi^2 M_p\ J_p} }\right)^{\frac{1}{2}}
\end{equation}
Here $M_p$ is mass of the pseudoscalar meson and $I_p$ and $J_p$ are defined as
\begin{equation}
I_p = \int_0^\infty dp \ p^2 A (p) [G_{q1} (p) G^*_{q2} (-p)]^{\frac{1}{2}}
\end{equation}
\begin{equation}
J_p = \int_0^\infty dp \ p^2  [G_{q1} (p) G^*_{q2} (-p)]
\end{equation}
respectively. Where,
\begin{equation}
A (p) =  \frac{(E_{p1}+m_{q1})(E_{p2}+m_{q2})-p^2}{[E_{p1}\ E_{p2} (E_{p1}+m_{q1})(E_{p2}+m_{q2})]^{\frac{1}{2}}}
\end{equation}
and $E_{p_i} = \sqrt{{k_i}^2 + m_{q_i}^2}$.

The computed decay constants of $D_s$ meson from $1S$ to $4S$ states are tabulated in Table \ref{tab8}. Present result for $1S$ state is compared with experimental as well as other model predictions. There are no model predictions available for comparison of the decay constants of the $2S$ to 4S states.

\begin{table*}
\begin{center}
\caption{Magnetic (M1) transition of Open Charm Meson}\label{tab9}
\begin{tabular}{cccccccccccccccc}
\hline\hline
 &  \multicolumn{3}{c} {k (MeV)} && \multicolumn{10}{c} {$\Gamma$ (keV)}\\
\cline{2-4} \cline{6-16}
Process & Present  &&\cite{Naynesh2011} && $ Present$  && PDG \cite{PDG2012} &&\cite{Naynesh2011}&&\cite{Jena1998} && \cite{HAKAN2001} && \cite{Radford2009} \\
\hline
(1S)$D^*_s \rightarrow D_s \gamma$ &  141.24&& 403 && 0.3443  && $<$ 4500&& 5.98&& 0.13  &&   0.48     && 1.12   \\
(2S)$D^*_s \rightarrow D_s \gamma$ &  83.48 && 152 && 0.0134 &&         && 0.35&&      &&            &&   \\
(3S)$D^*_s \rightarrow D_s \gamma$ &  61.21 && 91  && 0.0030 &&         && 0.08&&      &&            &&   \\
(3S)$D^*_s \rightarrow D_s \gamma$ &  49.47 && 65  && 0.0010 &&         && 0.03&&      &&            &&   \\
\hline\hline

\end{tabular}
\end{center}
\end{table*}

\begin{table*}
\begin{center}
\caption{Pseudoscalar  decay constant ($f_{P}$) of $D_s$ systems (in MeV).}\label{tab8}
\begin{tabular}{ccccccccccc}
\hline\hline
 &  \multicolumn{7}{c} {$f_{P}$}   & \\
\cline{2-8}
 & 1S &&  2S  && 3S &&  4S   \\
\hline
Present                            & 252.81		    && 336.56       &&391.74&&  433.16 \\
PDG \cite{PDG2012}                 &	260.0 $\pm$ 5.4	&	           &      &         \\
Belle  \cite{Zupanc2013}           & 255.5 $\pm$ 4.2 $\pm$ 5.1&     &      &  \\
BaBar \cite{PAmo2010}              &	258.6 $\pm$ 6.4 $\pm$ 7.5		&	           &      &         \\
CLEO-c \cite{Naik2009}             &	259.0 $\pm$ 6.2 $\pm$ 3.0		&	           &      &         \\
CLEO-c \cite{Onyisi2009}           &	252.6 $\pm$ 11.1 $\pm$ 5.2 		&	           &      &         \\
$[QCDSR]$ \cite{Narison2013}         &	246 $\pm$ 6		&	           &      &     \\
$[RPM]$ \cite{Mao2012}               & 256 $\pm$ 26       &	           &      &          \\
$[QCDSR]$ \cite{Lucha2011}           &	245.3 $\pm$ 15.7&	           &      &            \\
$[LQCD]$ \cite{Blossier2009}         & 244 $\pm$ 8	    &	           &      &          \\
$[LQCD]$ \cite{Davies2010,Na2012}    &248.0 $\pm$ 2.5  &	           &      &          \\
$[LQCD]$ \cite{Bazavov2012}          &260.1 $\pm$ 10.8 &	           &      &           \\
$[LFQM]$ \cite{Hwang2010}            & 264.5 $\pm$ 17.5&	           &      &         \\
$[QCDSR]$ \cite{Wang2013}            & 241 $\pm$ 12    &	           &      &            \\
$[RBSM]$ \cite{Wang}                 & 248 $\pm$ 27    &	           &      &         \\

\hline\hline

\end{tabular}
\begin{center}
[QCDSR]- QCD sum rule.\\
$[RPM]$- Relativistic potential Model.\\
$[LQCD]$- Lattice QCD.\\
$[LFQM]$- Light front quark model.\\
$[RBSM]$- Relativistic Bethe-Salpeter Method.\\
\end{center}
\end{center}
\end{table*}

\section{Leptonic Decay of the Open Heavy Flavour Mesons}
Charged mesons produced from a quark and anti-quark can decay to a charged lepton pair when
these objects annihilate via a virtual $W^\pm$ boson as given in Fig.(\ref{leptonic decay}). Though the leptonic decays of open flavour mesons belong to rare decay \cite{Hikasa1992,Rosner2008}, they have clear experimental signatures due to the presence of highly energetic lepton in the final state. And such decays are very clean due to the absence of hadrons in the final state \cite{Villa2007}. The leptonic width of $D_s$ meson is computed using the relation given by
\begin{eqnarray}\label{eq:af}
\Gamma(D^+_s \rightarrow l^+\nu_l)=\frac{G_F^2}{8\pi} f^2_{D_s} |V_{cs}|^2 m_l^2 \left(1-\frac{m_l^2}{M^2_{D_s}}\right)^2 M_{D_s}\ \ \ \ \
\end{eqnarray}
in complete analogy to $\pi^+ \rightarrow l^+ \nu$. These transitions are helicity suppressed; i.e., the amplitude is proportional to $m_l$, the mass of the lepton $l$. The leptonic widths of $D_s$ ($1^1S_0$ state) meson are obtained from Eqn.(\ref{eq:af}) where the predicted values of the pseudoscalar decay constant $f_{D_s}$ along with the masses of $M_{D_s}$ and the PDG value for $V_{cs}$ = 1.006 are used. The leptonic widths for separate lepton channel are computed for the choices of $m_{l=\tau, \mu, e}$. The branching ratio of these leptonic widths are then obtained as
\begin{equation}
BR = \Gamma (D_s \rightarrow l^+ \nu_l)\times \tau
\end{equation}
where $\tau$ is the experimental lifetime of the $D_s$ meson. The respective leptonic widths are tabulated in Table \ref{tab10} along with other model predictions as well as with the experimental values. Our results are found to be in
accordance with the available experimental values.



\begin{table*}
\begin{center}
\caption{The leptonic decay width and leptonic Branching Ratio (BR) of $D_s$ meson.}
\label{tab10}
\begin{tabular}{ccccccccccc}
\hline\hline
        &  \multicolumn{2}{c} {$\Gamma(M \rightarrow l \bar{\nu_l})$ (keV)}  &&  \multicolumn{7}{c} {BR  (keV)}   \\
        \cline{2-3}\cline{5-11}
Process &  Present              & \cite{HAKAN2000} && Present && \cite{Naynesh2011} && \cite{HAKAN2000} && Experiment \cite{PDG2012}  \\
\hline\\
$D_s \rightarrow \tau \bar{\nu_\tau}$  &7.508 $\times 10^{-8}$& 6.090 $\times 10^{-8}$ && 5.706 $\times 10^{-2}$&&4.22 $\times 10^{-2}$ && 4.3 $\times 10^{-2}$ &&           $(5.43 \pm 0.31 )\times 10^{-2}$ \\
$D_s \rightarrow \mu \bar{\nu_\mu}$    &7.648 $\times 10^{-9}$& 6.240 $\times 10^{-9}$ && 5.812 $\times 10^{-3}$&& 4.25 $\times 10^{-3}$ && 4.41 $\times 10^{-3}$ &&              $(5.90 \pm 0.33 ) \times 10^{-3}$\\
$D_s \rightarrow e \bar{\nu_e}$        &1.792 $\times 10^{-13}$&          . . .            && 1.362 $\times 10^{-7}$&& 1.00 $\times 10^{-7}$ &&       . . .    && $<1.2 \times 10^{-4}$\\
\hline\hline
\end{tabular}
\end{center}
\end{table*}


\section{Results and Discussion}
We have studied the mass spectra and decay properties of the $D_s$ meson in the framework of relativistic independent quark model. Our computed $D_s$ meson spectral states are in good agreement with the reported PDG values of known states. Though there are many excited $1^{-}$ state of $D_s$ meson known experimentally, most of them beyond 1S states are still not understood completely. And in the case of P-wave states only $1 ^3 P_{J} $, $1 ^1P_1$, and $2\ ^1P_1$ of the $D_s$ meson are known experimentally. Our results are also compared with other theoretical model predictions. \\

The predicted masses of S-wave $D_s$ meson state $2\ ^3S_1$ (2717.8 MeV) and $2\ ^1S_0$ (2633.5 MeV) are in very good agreement with experimental result of $2710^{+12}_{-7}$ MeV by BaBar \cite{Brodzicka2008} and Belle \cite{Aubert2006} Collaborations and 2638 MeV for $2\ ^1S_0$ by SELEX Collaboration \cite{Evdokimov}   respectively. The expected results of other S-wave excited states of $D_s$ meson are also in good agreement with other reported values  \cite{Badalian2011,Ebert2010,Naynesh2011,De2011}. The predicted P-wave $D_s$ meson states, $1^3P_2$ (2584.7 MeV), $1^3P_1$ (2516.9 MeV), $1^3P_0$ (2349.0 MeV) and $1^1P_1$ (2435.6 MeV) are in good agreement with experimental \cite{PDG2012} results of $2571.9 \pm 0.8 $ MeV, $2535.12 \pm 0.13 $ MeV, $2317.8 \pm 0.6 $ MeV and $2459.6 \pm 0.6 $ MeV respectively. The $2^3P_1$ (2986.4 MeV) and $1 ^3D_3$ (2931.4) are nearly 50-60 MeV off  with the experimental results of $3044^{+30}_{-9}$ MeV \cite{Aubert09} and $2862^{+6}_{-3}$ MeV \cite{Aubert09}. However their $J^P$ values are not yet confirmed experimentally. Though our predictions of $1S$, $2S$ and $1P$ states are in agreement with the experiment. The experimental state of $D_{sJ}^*$ (2860) is found to be a mixed states of $1 ^3D_3$ (2931.4) and $1 ^3D_1$ (2842.0) with a mixing probability given by $\cos^2 \theta  = 0.2013$ and that for $D_{sJ}$ (3040) is a mixed state of ($2 ^3P_2$ (3107.6) and $2 ^3P_0$ (2764.3) with a mixing probability given by $\cos^2 \theta  = 0.8030$.

In the relativistic Dirac formalism, the spin degeneracy is primarily broken therefore, to have spin average masses of the different spectral states we employ the spin averaging procedure as
\begin{equation}
M_{CW} = \frac{\sum_J (2 J +1) M_J}{\sum_J (2 J +1)}
\end{equation}
The spin average or the center of weight masses $M_{CW}$ are calculated from the known values of the different meson states and are compared with other model prediction \cite{Ebert2010} and those predicted by lattice QCD [LQCD] \cite{Mohler2011} in Table \ref{tab12}. It also help us to know the different spin dependent contributions for the observed state.

The precise experimental measurements of the masses of $D_s$ meson states provided a real test for the choice of the hyperfine and the fine structure interactions adopted in the study of $D_s$ meson spectroscopy. Recent study of  $D_s$ meson mass splittings in lattice QCD [LQCD] \cite{Mohler2011} using 2 $\pm$ 1 flavor configurations generated with the Clover-Wilson fermion action by the PACS-CS collaboration \cite{Mohler2011} has been used for comparison. Present results as seen in Table \ref{tab11} are in very good agreement with the respective experimental values over the lattice results \cite{Mohler2011}. In this Table, the present results on an average, are in agreement with the available experimental value within $6 \%$ of variations, while the lattice QCD predictions \cite{Mohler2011} show $20 \%$ of variations.\\

The magnetic transitions (M1) can probe the internal charge structure of hadrons, and therefore they will likely play an important role in determining the hadronic structures of $D_s$ meson. The present M1 transitions widths of $D_s$ meson states as listed in Table \ref{tab9} are in accordance with the model prediction of \cite{HAKAN2001} while the upper bound provided by PDG \cite{PDG2012} is very wide. We do not find any theoretical predictions for M1 transition width of excited states for comparison. Thus we only look forward to see future experimental support to our predictions. \\

The calculated pseudoscalar decay constant ($f_P$) of $D_s$ meson is listed in Table (\ref{tab8}) along with other model predictions as well as experimental results. The value of $f_{D_s} (1S)$ = 252.81 MeV obtained in our present study is in very good agreement with the experimental values provided by Belle \cite{Zupanc2013}, BaBar \cite{PAmo2010} and CLEO-c \cite{Naik2009,Onyisi2009}. The present value is also in accordance with other theoretical predictions for $1S$ state.  The predicted $f_{D_s}$ for higher S-wave states are found to increase with energy. However, there are no experimental or theoretical values available for comparison.  Another important property of $D_s$ meson studied in the present case is the leptonic decay widths. The present branching ratios for $D_s \rightarrow \tau \bar{\nu_\tau}$ ($5.706 \times 10^{-2}$) and $D_s \rightarrow \mu \bar{\nu_\mu}$ ($5.812 \times 10^{-3}$) are in excellent agreement with the experimental results $ (5.43 \pm 0.31) \times 10^{-2}$ and  $(5.90 \pm 0.33) \times 10^{-3}$ respectively over other theoretical predictions vide Table \ref{tab10}. Large experimental uncertainty in the electron channel make it difficult for any reasonable conclusion.\\

Finally we look forward to see future high luminosity improved statistics and higher confidence level experimental data in support of our prediction on the spectroscopy and decay properties of the open charm-strange meson.

\section*{Acknowledgments}
The work is part of Major research project NO. F. 40-457/2011(SR) funded by UGC, INDIA.
One of the authors (Bhavin Patel) acknowledges the support through Fast Track project funded by DST (SR/FTP/PS-52/2011).

\end{document}